\documentclass[12pt]{iopart}

\usepackage{iopams,graphicx,amssymb}
\eqnobysec

\begin{document}


\centerline{EFFECTS OF TURBULENT TRANSFER ON THE CRITICAL BEHAVIOUR}

\centerline{N.\,V. Antonov, A.\,S. Kapustin, A.\,V. Malyshev}


\begin{abstract}
Critical behaviour of two systems, subjected to the turbulent mixing,
is studied by means of the field theoretic renormalization group. The
first system, described by the equilibrium model {\it A}, corresponds
to relaxational dynamics of a non-conserved order parameter. The second
one is the strongly nonequilibrium reaction-diffusion system, known as
Gribov process or directed percolation process. The turbulent mixing is
modelled by the stochastic Navier-Stokes equation with random stirring
force with the correlator $\propto \delta(t-t') p^{4-d-y}$, where $p$
is the wave number, $d$ is the space dimension and $y$ the arbitrary
exponent. It is shown that, depending on the relation between $y$ and
$d$, the systems exhibit various types of critical behaviour. In addition
to known regimes (original systems without mixing and passively advected
scalar field), existence of new strongly nonequilibrium universality classes
is established, and the corresponding critical dimensions are calculated to
the first order of the double expansion in $y$ and $\varepsilon=4-d$
(one-loop approximation).
\end{abstract}

\section{Introduction} \label{sec:Intro}

Various systems of very different physical nature exhibit interesting
singular behaviour in the vicinity of their critical points. Their
correlation functions acquire self-similar form with universal
critical dimensions: they depend only on few global characteristics
of the system (like symmetry or space dimension). Quantitative description
of critical behaviour is provided by the field theoretic renormalization
group (RG). In the
RG approach, possible types of critical regimes (universality classes) are
associated with infrared (IR) attractive fixed points of renormalizable field
theoretic models. Most typical equilibrium phase transitions belong to the
universality class of the $O_{n}$-symmetric $\psi^{4}$ model of an
$n$-component scalar order parameter. Universal characteristics of the
critical behaviour depend only on $n$ and the space dimension $d$ and can
be calculated within various systematic perturbation schemes, in particular,
in the form of expansions in $\varepsilon=4-d$ or $1/n$; see the monographs
\cite{Zinn,Book3} and the literature cited therein.

Aleksandr Nikolaevich Vasiliev made valuable contribution to the development
of field theoretic methods and their application in the theory of critical
behaviour and theory of turbulence. His work in this field is summarized in
the three monographs \cite{Book3,funk,turbo}. The most remarkable specific
achievements are probably the calculation of Fisher's exponent $\eta$ in the
$O_{n}$-symmetric $\psi^{4}$ model to the order $1/n^{3}$ \cite{n3} and the
third-order calculation of the anomalous exponents in Kraichnan's model of
turbulent advection \cite{Kr3}. In the present paper, we apply the field
theoretic RG and generalized $\varepsilon$ expansion to the problem of the
effects of turbulent transfer on various types of critical behaviour.

Over the past few decades, constant interest has been attracted by the
spreading processes and corresponding nonequilibrium phase transitions;
see e.g. the review papers \cite{Hinr,JT} and the literature cited therein.
Spreading processes are encountered in physical, chemical, biological and
ecological systems: autocatalytic reactions, percolation in porous media,
epidemic diseases and so on.
The transitions between the fluctuating (active) and absorbing (inactive)
phases, where all the fluctuations cease entirely, are
especially interesting as examples of nonequilibrium critical behaviour.

It has long been realized that the behaviour of a real critical system is
extremely sensitive to external disturbances, gravity, impurities and
turbulent mixing; see the monograph \cite{Ivanov} for the general discussion
and references. What is more, some disturbances (randomly distributed
impurities or turbulent mixing) can produce completely new types of critical
behaviour with rich and rather exotic properties.

These issues become even more important for nonequilibrium phase
transitions, because the ideal conditions of a ``pure'' stationary critical
state can hardly be achieved in real chemical or biological systems, and
the effects of various disturbances can never be completely excluded.
In particular, intrinsic turbulence effects cannot be avoided in chemical
catalytic reactions or forest fires. One can also speculate that atmospheric
turbulence can play important role for the spreading of an infectious disease
by flying insects or birds. Effects of different kinds of regular and
turbulent flows on the critical behaviour were studied in
\cite{Onuki}--\cite{AIK}.

In a number of papers \cite{Satten}--\cite{AIK}, critical behaviour of
various systems, subjected to the turbulent mixing, was studied by
means of the field theoretic RG. As a rule, the turbulence was modelled
by the time-decorrelated Gaussian velocity field
with the velocity correlation function of the form
$\langle vv\rangle \propto \delta(t-t') \, p^{-d-\xi}$, where
$p$ is the wave number and $0<\xi<2$ is a free parameter with the
real (``Kolmogorov'') value $\xi=4/3$.
This ``Kraichnan's rapid-change model'' has attracted enormous
attention recently because of the insight
it offers into the origin of intermittency and anomalous scaling in
fully developed turbulence; see the review paper \cite{FGV} and references
therein. The RG approach to that problem is reviewed in \cite{JPhysA}.
In the context of our study it is especially important that
Kraichnan's ensemble allows one to easily model anisotropy of
the flow \cite{Alexa} and compressibility of the fluid \cite{AIK},
which appears much more difficult if the velocity is described
by the full-scale dynamical equations.

However, the Gaussianity and vanishing correlation time are drastic
simplifications of the real situation, and it is desirable to investigate
effects of turbulent mixing, caused by more realistic velocity fields.
In this paper, we study effects of a strongly non-Gaussian velocity field
with finite correlation time, governed by a stochastic dynamical equation.
More precisely, we employ the stochastic
Navier--Stokes equation for an incompressible velocity, with random stirring
force with the correlator $\propto p^{4-d-y}$, where $y$ is the arbitrary
exponent with the physical (``Kolmogorov'')  value  $y=4$. The RG approach
to this model is reviewed in \cite{Book3,turbo}.

Two representative cases of dynamical critical behaviour are considered:
equilibrium  critical dynamics of a non-conserved order parameter with
$\psi^4$-type Hamiltonian, and the nonequilibrium system near its
transition point between the absorbing and fluctuating states. The
former model corresponds to critical fluid systems (binary mixtures or
liquid crystals), and the latter describes the spreading processes in
reaction-diffusion  systems, belongs to the  universality class known
as Gribov process or directed percolation process, and is equivalent
(up to the Wick rotation) to the Reggeon field theory \cite{Hinr,JT}.

It is shown that, depending on the relation between $y$ and $d$, the both
systems exhibit various types of critical  behaviour,  associated  with
different IR attractive fixed points of the RG equations. In addition to
known asymptotic regimes (like equilibrium critical dynamics without mixing
or passively advected scalar without self-interaction), existence of new,
strongly nonequilibrium, types of critical behaviour (universality
classes) is established, and the corresponding domains of stability in the
$y$--$d$ plane and the critical dimensions are calculated to the leading
order of the double expansion in $y$ and $\varepsilon=4-d$, which
corresponds to the one-loop approximation of the RG.

\section{Description of the models} \label{sec:Models}

In the Langevin formulation the models are defined by stochastic differential
equations for the order parameter $\psi = \psi(t,{\bf x})$:
\begin{eqnarray}
\partial_{t} \psi = \lambda_{0} \left\{ (-\tau_{0} +
\partial^{2}) \psi - u_{0} \psi^{3}/3! \right\} + \zeta
\label{stohA}
\end{eqnarray}
for the model {\it A} and
\begin{eqnarray}
\partial_{t} \psi = \lambda_{0} \left\{ (-\tau_{0} +
\partial^{2}) \psi - g_{0} \psi^{2}/2 \right\} + \zeta \sqrt{\psi}
\label{stohG}
\end{eqnarray}
for the Gribov process. Here, $\partial_{t}= \partial/ \partial t$,
$\partial^{2}$ is the Laplace operator, $g_{0}$ and $u_{0}>0$ are
the coupling constants, $\lambda_{0}>0$ is the kinematic (diffusion)
coefficient and $\tau_{0} \propto (T-T_{c})$ is the deviation of the
temperature (or its analog) from the critical value. The Gaussian random
noise $\zeta=\zeta(t,{\bf x})$ with zero average is specified by the pair
correlation function:
\begin{eqnarray}
\langle \zeta (t,{\bf x})\zeta (t',{\bf x'}) \rangle =
2 \lambda_{0}  \delta(t-t')\delta^{(d)}({\bf x}-{\bf x}')
\label{kor1}
\end{eqnarray}
for the model {\it A} and
\begin{equation}
\langle \zeta (t,{\bf x})\zeta (t',{\bf x'}) \rangle = g_{0}\lambda_{0}\,
\delta(t-t')\delta^{(d)}({\bf x}-{\bf x}')
\label{shum}
\end{equation}
for the Gribov process; $d$ being the dimension of the ${\bf x}$ space.
The factor $\sqrt\psi$ in the noise term of (\ref{stohG}) guarantees that
in the absorbing state the fluctuations cease entirely. The expressions
for the correlators differ only by normalization: the factor $2\lambda_{0}$
in (\ref{kor1}) is dictated by the fluctuation-dissipation relation and
ensures the correspondence to the static $\psi^{4}$ model, while
$g_{0}\lambda_{0}$ in (\ref{shum}) provides the simple form of the symmetry
that exists in the field theoretic formulation of the Gribov model;
see eq. (\ref{symm}) below. The subscript ``0'' marks the bare
(unrenormalized) parameters; their renormalized analogs (without the
subscript) will appear later.

For incompressible fluid, the Galilean covariant coupling with the
transverse (due to the incompressibility condition $\partial_i v_i=0$)
velocity field ${\bf v}= \{ v_{i}(t, {\bf x}) \}$ is introduced by
the substitution
\begin{eqnarray}
\partial_{t} \to \nabla_{t} = \partial_{t} + v_{i} \partial_{i}
\label{nabla}
\end{eqnarray}
in (\ref{stohA}) and (\ref{stohG}), where
$\partial_i = \partial /\partial x_{i}$ and $\nabla_{t}$ is the
Lagrangian (Galilean covariant) derivative. We will employ the velocity
field satisfying the NS equation with a random stirring force
\begin{equation}
\nabla _t v_k=\nu _0\partial^{2} v_k - \partial_k {\cal P}+f_k,
\label{1.1}
\end{equation}
where $\nabla _t$ is the Lagrangian derivative (\ref{nabla}),
${\cal P}$ and $f_k$ are the pressure
and the transverse random force per unit mass. We assume for $f$
a Gaussian distribution with zero average and correlation function
\begin{equation}
\big\langle f_i(x)f_j(x')\big\rangle = \frac{\delta (t-t')}{(2\pi)^{d}}\,
\int_{p\ge m} d{\bf p}\, P_{ij}({\bf p})\, {\cal D}_f(p)\, \exp
\left\{{\rm i}{\bf p} \left({\bf x}-{\bf x}'\right)\right\} ,
\label{1.2}
\end{equation}
where $P_{ij}({\bf p}) =\delta _{ij}  - p_i p_j / p^2$ is the transverse
projector and ${\cal D}_f(p)$ is some function of $p= |{\bf p}|$ and model
parameters. The momentum $m=1/{\cal L}$, the reciprocal of the integral
turbulence scale ${\cal L}$, provides IR regularization (its precise form
is unessential;
the sharp cutoff is the simplest choice for the practical calculations).

The standard RG formalism is applicable to the problem (\ref{1.1}),
(\ref{1.2}) if the correlation function of the random force is chosen
in the power form
\begin{equation}
{\cal D}_f(p)=D_0\,p^{4-d-y},
\label{1.9}
\end{equation}
where $D_{0}>0$ is the positive amplitude factor and the exponent
$0<y\le 4$ plays the role of the RG expansion parameter, analogous
to that played by $\varepsilon=4-d$ in models of critical behaviour.
Its physical value is $y=4$: with the appropriate choice of the amplitude,
the function (\ref{1.9}) for $y\to4$ turns to the $\delta$ function,
${\cal D}_f(p) \propto \delta({\bf p})$, which corresponds to the injection
of energy to the system owing to interaction with the largest
turbulent eddies; for a more detailed discussion of this point
see e.g.~\cite{Book3,turbo}.

\section{Field theoretic formulation and renormalization} \label{sec:QFT}

The stochastic problems (\ref{stohA})--(\ref{shum}) can be reformulated
as field theoretic models of the doubled set of fields
$\Psi = \{\psi,\psi^{\dag}\}$ with action functional
\begin{eqnarray}
{\cal S}_{A} (\psi,\psi^{\dag}) =  \psi^{\dag}
\left(-\partial_{t}+\lambda_{0} \partial^{2}- \lambda_{0}\tau_{0}\right)
\psi + \lambda_{0} \psi\psi^{\dagger} - \frac{u_{0}}{3!} \psi^{\dag}\psi^3
\label{actionA}
\end{eqnarray}
for the model {\it A} and
\begin{eqnarray}
{\cal S}_{G} (\psi,\psi^{\dag}) =  \psi^{\dag}
(-\partial_{t}+\lambda_{0} \partial^{2}- \lambda_{0}\tau_{0}) \psi
+ \frac{g_{0}\lambda_{0}}{2} \left\{ (\psi^{\dagger})^2\psi -
\psi^{\dagger}\psi^2  \right\}
\label{actionG}
\end{eqnarray}
for the Gribov model.
Here, $\psi^{\dag}=\psi^{\dag}(t,{\bf x})$ is the auxiliary ``response
field'' and the integrations over the arguments of the fields are implied,
for example
\[  \psi^{\dag}\partial_{t}\psi = \int dt \int d{\bf x}
\psi^{\dag}(t,{\bf x})\partial_{t}\psi(t,{\bf x}). \]

The stochastic problem (\ref{1.1})--(\ref{1.9}) corresponds to the
field theoretic model with the action
\begin{equation}
{\cal S}_{N\!S}({v}', {v}) =
v 'D_{v}v'/2+ v'\left\{-\nabla_t +\nu _0\partial^{2} \right\} v,
\label{actionV}
\end{equation}
where $D_{v}$ is the correlation function (\ref{1.2}) and all the needed
integrations and summations over the vector indices are understood.
The auxiliary vector field ${\bf v}'= \{ v'_{i}(t,{\bf x})\}$ is also
transverse, $\partial_{i}v_{i}'=0$, which allows one to omit the pressure
term in the action functional (\ref{actionV}).

The field theoretic formulation means that statistical averages
of random quantities in the original stochastic problems can be represented
as functional integrals over the full set of fields with the weight
$\exp {\cal S}(\Phi)$, and can therefore be viewed as the Green functions
of the field theoretic models with actions (\ref{actionA})--(\ref{actionF}).
In particular, the linear response function of the stochastic problems
(\ref{stohA})--(\ref{shum}) is given by the Green function
\begin{eqnarray}
G=\langle \psi^{\dag}(t, {\bf x}) \psi(t', {\bf x'} ) \rangle =
\int {\cal D}\psi^{\dag} \int {\cal D} \psi\ \,
\psi^{\dag}(t, {\bf x}) \psi(t', {\bf x'})\, \exp {\cal S}(\psi,\psi^{\dag})
\label{respd}
\end{eqnarray}
of the corresponding field theoretic models.

The model (\ref{actionG}) is symmetric with respect to the transformation
\begin{equation}
\psi(t,{\bf x})\to \psi^{\dagger}(-t,-{\bf x}),
\quad \psi^{\dagger}(t,{\bf x})\to \psi(-t,-{\bf x}),
\quad g_{0}\to -g_{0}.
\label{symm}
\end{equation}
Reflection of the constant $g_{0}$ is in fact unimportant because
the actual expansion parameter in the perturbation theory of the model
is $u_{0}= g_{0}^{2}$. The model (\ref{actionA}) is symmetric with
respect to the reflection of the fields $\psi\to-\psi$,
$\psi^{\dagger}\to-\psi^{\dagger}$. These symmetries survive
the inclusion of the velocity field.

The full-scale models are described by the action functionals
\begin{equation}
{\cal S}_{A,G}^{F} (\Psi) = {\cal S}_{N\!S} (v',v) +
{\cal S}_{A,G} (\psi,\psi^{\dag},v),
\label{actionF}
\end{equation}
where  $\Psi = \{\psi,\psi^{\dag},v,v'\}$ is the full set of fields and
the substitution (\ref{nabla}) is made in the functionals (\ref{actionA})
and (\ref{actionG}). For these models, the full set of coupling constants
(``charges'') involves the three parameters
\begin{equation}
u_{0}  \sim \Lambda^{4-d}, \quad w_{0} = D_{0}/\nu_{0}^{3} \sim \Lambda^{y},
\quad e_{0} = \lambda_{0}/\nu_{0},
\label{charges}
\end{equation}
where $\Lambda$ is some typical UV momentum scale. The ratio $e_{0}$ is not
an expansion parameter in the perturbation theory, but it should also be
treated as an additional coupling constant because it is dimensionless and
the renormalization constants and RG functions depend on it.

From the relations (\ref{charges}) it follows that the interactions
$(\psi^{\dagger})^2\psi$ and $\psi^{\dagger}\psi^2$ in (\ref{actionG})
and $\psi^{\dagger}\psi^{3}$ in (\ref{actionA}) become logarithmic (the
corresponding coupling constant $u_{0}$ becomes dimensionless) at $d=4$.
Thus for the single-charge problems (\ref{actionA}), (\ref{actionG}),
the value $d=4$ is the upper critical dimension, and the deviation
$\varepsilon=4-d$ plays the part of the formal expansion parameter in
the RG approach: the critical exponents are nontrivial for $\varepsilon>0$
and can be calculated as series in $\varepsilon$. The vertex term
$v'(v\partial)v$ in (\ref{actionV}) and the additional interactions
$\psi^{\dag}(v\partial)\psi$ in the full models (\ref{actionF}) become
logarithmic at $y=0$. The parameter $y$ is not related to $d$ and can
be varied independently. However, for the RG analysis of the full problems
it is important that all the interactions become logarithmic at the same
time. Otherwise, one of them would be weaker than the others from the RG
viewpoints and it would be irrelevant in the leading-order IR behaviour.
As a result, some of the scaling regimes of the full model would be lost.

In order to study all possible scaling regimes and the crossovers between
them, we need a genuine three-charge theory, in which all the interactions
are treated on equal footing. Thus we will treat $\varepsilon$ and $y$
as small parameters of the same order, $\varepsilon \propto y$.
Instead of the plain $\varepsilon$ expansion in the single-charge models,
the coordinates of the fixed points, critical dimensions
and other quantities will be calculated as double expansions in the
$\varepsilon$--$y$ plane around the origin, that is, around the point
in which all the coupling constants in (\ref{charges}) become
dimensionless.

The analysis based on the dimensionality considerations and the symmetries
of the full-scale models (\ref{actionF}) shows that they are multiplicatively
renormalizable. The role played by the symmetries is very important:
in particular, the Galilean invariance requires that the counterterms
$\psi^{\dag} \partial _{t}\psi$ and $\psi^{\dag} (v\partial) \psi$
enter the renormalized action only in the form of invariant combination
$\psi^{\dag} \nabla _{t}\psi$. It also shows that the counterterm
$\psi^{\dag} \psi v^{2}$, absent in the unrenormalized actions
(\ref{actionF}) and allowed by the dimension, is in fact forbidden.
Thus, all the UV divergences (having the form of singularities at
$\varepsilon$ and $y\to0$) can be absorbed into a finite set of
renormalization constants $Z_{i}$. The renormalized action functionals
have the forms:
\begin{equation}
{\cal S}_{A}^{R} = {\cal S}_{N\!S}^{R} +
\psi^{\dag}
\left\{ -Z_{1}\partial_{t}+Z_{2}\lambda \partial^{2}
- Z_{3}\lambda\tau\right\}\psi + Z_{4}\lambda \psi\psi^{\dagger}
-Z_{5} \frac{u\mu^{\varepsilon} \lambda}{3!} \psi^{\dag}\psi^3
\label{actionAR}
\end{equation}
for the model {\it A} and
\begin{equation}
{\cal S}_{G}^{R} = {\cal S}_{N\!S}^{R} +
\psi^{\dag}
\left\{-Z_{1}\partial_{t}+Z_{2}\lambda \partial^{2}- Z_{3}\lambda\tau\right\}
\psi +Z_{4}
\frac{g\mu^{\varepsilon/2}\lambda}{2} \left\{ (\psi^{\dagger})^2\psi -
\psi^{\dagger}\psi^2  \right\}
\label{actionGR}
\end{equation}
for the Gribov model, where ${\cal S}_{N\!S}^{R}$ is obtained from
(\ref{actionV}) by the substitution $\nu_0\to\nu Z_{\nu}$ and
$w_{0} \nu_0^{3}\to w \mu^y\nu^{3}$ (the nonlocal term with the random force
correlator in (\ref{actionV}) is not renormalized). Here and below
$\tau$, $u$ and so on are renormalized analogs of the bare parameters
(with the subscripts ``0'') and $\mu$ is the reference mass
scale (additional arbitrary parameter of the renormalized theory).

\begin{figure}
\begin{center}
\includegraphics[width=12cm]{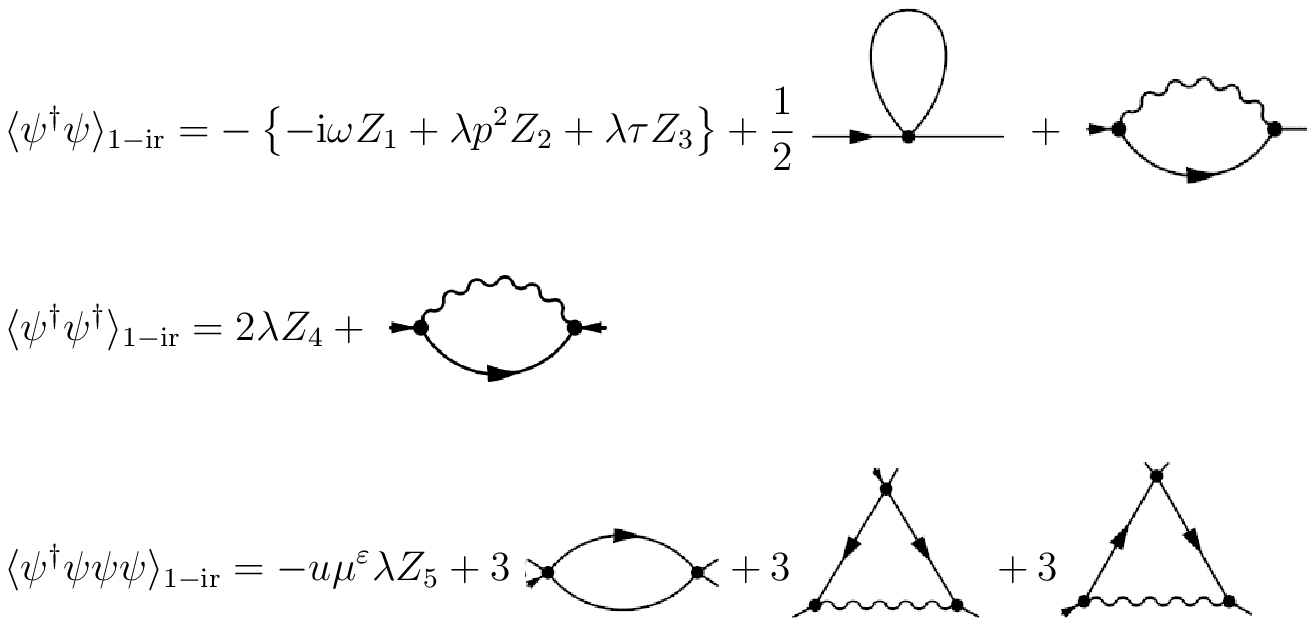}
\caption{\label{fig:A}
One-loop approximation for the relevant 1-irreducible Green functions
in the model (\protect\ref{actionAR}).}
\end{center}
\end{figure}

The one-loop calculation of the renormalization constants $Z_{i}$ is easily
performed: in fact, in this approximation there are no new Feynman diagrams
in comparison to the models (\ref{actionA})--(\ref{actionV})
and the passive scalar case. More precisely, all the new diagrams appear UV
finite for an incompressible fluid and give no contribution to $Z_{i}$.
This fact is illustrated for the model {\it A} in figure~\ref{fig:A}, where
all 1-irreducible Green functions, needed for the calculation of the
renormalization constants, are shown in the one-loop approximation.
The solid lines with arrows denote the bare propagator
$\langle \psi\psi^{\dag} \rangle_{0}$, the arrow pointing to the field
$\psi^{\dag}$. The solid lines without arrows correspond to the propagator
$\langle \psi\psi \rangle_{0}$ and the wavy lines denote the velocity
propagator $\langle vv \rangle_{0}$. The external ends with incoming arrows
correspond to the fields $\psi^{\dag}$, the ends without arrows correspond
to $\psi$. The quartic vertex with one incoming arrow corresponds to the
interaction $- {u\mu^{\varepsilon} \lambda} \psi^{\dag}\psi^3/{3!}$, while
the triple vertex with one wavy line corresponds to
$-\psi^{\dag}(v\partial)\psi$. Due to the transversality of the velocity
field, the derivative at the latter vertex can also be moved onto the field
$\psi^{\dag}$ using integration by parts:
$-\psi^{\dag} (v\partial) \psi = \psi (v\partial) \psi^{\dag}$. Thus
in any diagram involving $n$ external vertices of this type, the factor
$p^{n}$ with $n$ external momenta $p$ will be taken outside the
corresponding integrals. This reduces the dimension of the integrand by
$n$ units and can make it UV convergent. In the case at hand, this proves
the UV finiteness of the last two diagrams in the function
$\langle \psi^{\dag}\psi\psi\psi \rangle_{1-ir}$ and the only diagram in
$\langle \psi^{\dag}\psi^{\dag} \rangle_{1-ir}$ (which otherwise would be
logarithmically divergent). Since in our models the scalar field is passive
(no feedback on the velocity statistics), the constant $Z_{\nu}$ in the
full model (\ref{actionF}) is the same as in (\ref{actionV}), but with
the substitution $d=4$.

In the minimal subtraction scheme the one-loop expressions for the
constants $Z_{i}$ contain only simple poles in $\varepsilon$ and $y$
and have the forms:
\begin{eqnarray}
Z_{1} = Z_{4} = 1, \quad
Z_{3} = 1 + \frac{u}{\varepsilon}, \quad
Z_{2} = 1 - \frac{w}{4ye(e+1)} , \quad
Z_{5} = 1 + \frac{3u}{\varepsilon}
\label{ZoA}
\end{eqnarray}
for the model {\it A},
\begin{eqnarray}
Z_{1} = 1 + \frac{u}{4\varepsilon}, \quad
Z_{3} = 1 + \frac{u}{2\varepsilon}, \quad
Z_{4} = 1 + \frac{u}{\varepsilon},
\nonumber \\
Z_{2} = 1 + \frac{u}{8\varepsilon}- \frac{w}{4ye(e+1)}
\label{ZoG}
\end{eqnarray}
for the Gribov model and
\begin{eqnarray}
Z_{\nu} = 1 - \frac{w}{12y}
\label{ZoV}
\end{eqnarray}
for the both cases (in order to simplify the coefficients, the factor
$1/16\pi^2$ is absorbed into the constants $u$ and $w$).

\section{Fixed points and scaling regimes} \label{sec:FPS}

The RG equations for our multiplicatively renormalized models
(\ref{actionAR}), (\ref{actionGR}) are derived in a standard fashion,
similar to that for the analogous models with Kraichnan's velocity field
\cite{AIK}, and we do not present them here.

It is well known that possible IR scaling regimes of a renormalizable field
theoretic model are associated with IR attractive fixed points of the
corresponding RG equations; see e.g. \cite{Zinn,Book3}. For a given point,
the Green functions demonstrate self-similar (scaling) asymptotic behaviour
in the IR range, with definite critical dimensions $\Delta_{F}$ of all fields
and parameters $F$ of the model. The coordinates $g_{i*}$ of the fixed points
are found from the requirement that the $\beta$-functions, corresponding to
all renormalized couplings $g_{i}$, vanish. The type of a fixed point is
determined by the matrix $\Omega_{ik}=\partial\beta_{i}/\partial g_{k}$,
where $\beta_{i}$ is the full set of $\beta$-functions and $g_{k}$ the
full set of couplings. For an IR attractive fixed point the matrix $\Omega$
is positive, i.e., the real parts of all its eigenvalues are positive.

In our case, $g_{i}= \{u,w,e\}$. Admissible fixed point must be IR
attractive for some values of $y$ and $\varepsilon$ and satisfy the
conditions $u_{*}, w_{*}, e_{*}>0$, which follow from the physical
meaning of these parameters. The functions $\beta_{i}$, calculated in
the one-loop approximation from the renormalization constants
(\ref{ZoA})--(\ref{ZoV}), have the forms:
\begin{equation}
\beta_{u} = u \left\{-\varepsilon + 3u +\frac{w}{2e(e+1)} \right\},
\quad \beta_{e} = w \left\{ \frac{e}{12} +\frac{w}{4(e+1)} \right\}
\label{betaA}
\end{equation}
for the model {\it A},
\begin{equation}
\beta_{u} = u \left\{-\varepsilon + \frac{3u}{2} +\frac{w}{2e(e+1)}
\right\},
\quad \beta_{e} = w \left\{ \frac{e}{12}- \frac{w}{4(e+1)} \right\}
- \frac{ue}{8}
\label{betaG}
\end{equation}
for the Gribov model and
\begin{equation}
\beta_{w} = w \left\{-y+w/4\right\}
\label{betag}
\end{equation}
for the both models, with higher-order corrections in $u$ and $w$.

The analysis
of the functions (\ref{betaA}), (\ref{betag}) reveals four admissible
fixed points of the model {\it A}:

(1) The Gaussian (free) fixed point: $u_{*}=w_{*}=0$, $e_{*}$ arbitrary.
This point is IR attractive for $y<0$, $\varepsilon<0$. The critical
dimensions are found exactly:
\[ \Delta_{\psi}=d/2-1, \quad  \Delta_{\psi^{\dag}}=d/2+1, \quad
\Delta_{\omega}=\Delta_{\tau}=2. \]

(2) The point $u_{*}=0$,  $w_{*}=4y$,  $2e_{*} = -1+\sqrt{13}$ (the positive
root of the equation $e(e+1)=3$), corresponding to the passively advected
scalar without self-interaction: the vertex $\psi^{\dag}\psi^{3}$ in
(\ref{actionA}) is IR irrelevant in the sense of Wilson.
This point is IR attractive for $y>0$, $y>3\varepsilon/2$.
The critical dimensions are also known exactly:
\[ \Delta_{\psi}=d/2-1, \quad  \Delta_{\psi^{\dag}}=d/2+1, \quad
\Delta_{\omega}=\Delta_{\tau}=2-y/3. \]

(3) The point $w_{*}=0$, $u_{*}=\varepsilon/3$, $e_{*}$ arbitrary,
corresponding to the pure {\it A} model: the turbulent advection is
IR irrelevant.\footnote{This becomes obvious if,
by rescaling the fields, the coupling constant $w_{0}$ is placed in
front of the interaction terms $\psi'(v \partial) \psi$, which is more
familiar for the field theory. We do not do it, however, in order to
retain the natural form of the covariant derivative (\protect\ref{nabla}).}
This point is IR attractive for $y<0$, $\varepsilon>0$. The critical
dimensions for this regime depend only on $\varepsilon$:
\[ \Delta_{\psi}=1-\varepsilon/2, \quad
\Delta_{\psi^{\dag}}=3-\varepsilon/2, \quad
\Delta_{\omega}=2, \quad \Delta_{\tau}=2-\varepsilon/3, \]
with the higher-order corrections, known up to $\varepsilon^{4}$ for
$\Delta_{\omega,\psi^{\dag}}$ \cite{Levca} and $\varepsilon^{5}$ for
the others \cite{Zinn,Book3}.

(4) The most interesting point $w_{*}=4y$, $u_{*}=\varepsilon/3-2y/9$,
$2e_{*}=-1+\sqrt{13}$, IR attractive for $y>0$, $y<3\varepsilon/2$. It
corresponds to a new full-scale nonequilibrium universality class, in
which both the self-interaction and turbulent mixing are relevant.
Here, the critical dimensions are calculated in the form of double series
in $\varepsilon$ and $y$. The one-loop expressions read:
\begin{equation}
\Delta_{\psi}=1- \frac{4(\varepsilon+y)}{3}, \quad
\Delta_{\psi^{\dag}}= 3 - \frac{4(\varepsilon+y)}{3}, \quad
\Delta_{\tau}=2-\varepsilon+ \frac{y}{3}, \quad
\Delta_{\omega}=2-\frac{y}{3}.
\label{dim4A}
\end{equation}
The last dimension is exact, the others have higher-order corrections
in $\varepsilon$ and $y$.

For the Gribov case, the analysis of the functions (\ref{betaG}),
(\ref{betag}) reveals {\it five} admissible fixed points:

(1) The Gaussian point: $u_{*}=w_{*}=0$, $e_{*}$ arbitrary, attractive
for $y<0$, $\varepsilon<0$. Here, the critical dimensions are:
\[ \Delta_{\psi}= \Delta_{\psi^{\dag}}=d/2, \quad
\Delta_{\omega}=\Delta_{\tau}=2. \]

(2) The point $u_{*}=0$,  $w_{*}=4y$,  $2e_{*} = -1+\sqrt{13}$,
attractive for $y>0$, $y>3\varepsilon/2$. It corresponds
to the passively advected scalar without self-interaction: the vertices
$(\psi^{\dagger})^2\psi$ and $\psi^{\dagger}\psi^2$ in (\ref{actionG})
are irrelevant. The critical dimensions are:
\[ \Delta_{\psi}= \Delta_{\psi^{\dag}}=d/2, \quad
\Delta_{\omega}=\Delta_{\tau}=2-y/3. \]

(3) The point $w_{*}=0$, $u_{*}=2\varepsilon/3$, $e_{*}=\infty$, IR
attractive for $y<0$, $\varepsilon>0$. It corresponds to the pure Gribov
process (turbulent advection is irrelevant). The critical
dimensions depend only on $\varepsilon$:
\begin{equation}
\Delta_{\psi}= \Delta_{\psi^{\dag}}= 2- 7\varepsilon/12, \quad
\Delta_{\tau}=2-\varepsilon/4, \quad
\Delta_{\omega}=2- \varepsilon/12,
\label{dim2G}
\end{equation}
with known corrections of order $\varepsilon^{2}$ \cite{JT}.

(4) The full-scale point, corresponding to a new universality class:
$w_{*}=4y$, $u_{*}=4\varepsilon/5-8y/15$,
$2e_{*}=-1+\sqrt{1+40y/(4y-\varepsilon)}$.
This point is IR attractive for $y>\varepsilon/4$, $y<3\varepsilon/2$.
The dimensions are calculated as double series in $\varepsilon$ and $y$
with the one-loop expressions:
\begin{equation}
\Delta_{\psi}=\Delta_{\psi^{\dag}}=2-\frac{3\varepsilon}{5}+\frac{y}{15},
\quad
\Delta_{\tau}=2- \frac{y+\varepsilon}{5}, \quad
\Delta_{\omega}=2-\frac{y}{3}\ {\rm (exact)} .
\label{dim4G}
\end{equation}

(5) The point $u_{*}=2\varepsilon/3$, $w_{*}=4y$, $e_{*} = \infty$,
IR attractive for $y>\varepsilon/4$, $\varepsilon>0$. This point requires
a careful interpretation. Although the value of $w_{*}$ at this point is
nontrivial (and the velocity field is therefore non-Gaussian), the
turbulent mixing is nevertheless irrelevant. Indeed, straightforward
analysis of the Green functions with the scalar fields $\psi$, $\psi^{\dag}$,
for example the function (\ref{respd}), shows that the Feynman diagrams
involving the velocity field vanish in the limit $e = \lambda/\nu \to\infty$,
while the diagrams without the velocity are independent of $e$ and remain
finite. Thus from the physics viewpoints this regime is similar to (3)
and corresponds to the pure Gribov process, and the corresponding critical
dimensions indeed coincide with (\ref{dim2G}).

\begin{figure}
\begin{center}
\includegraphics[width=11cm]{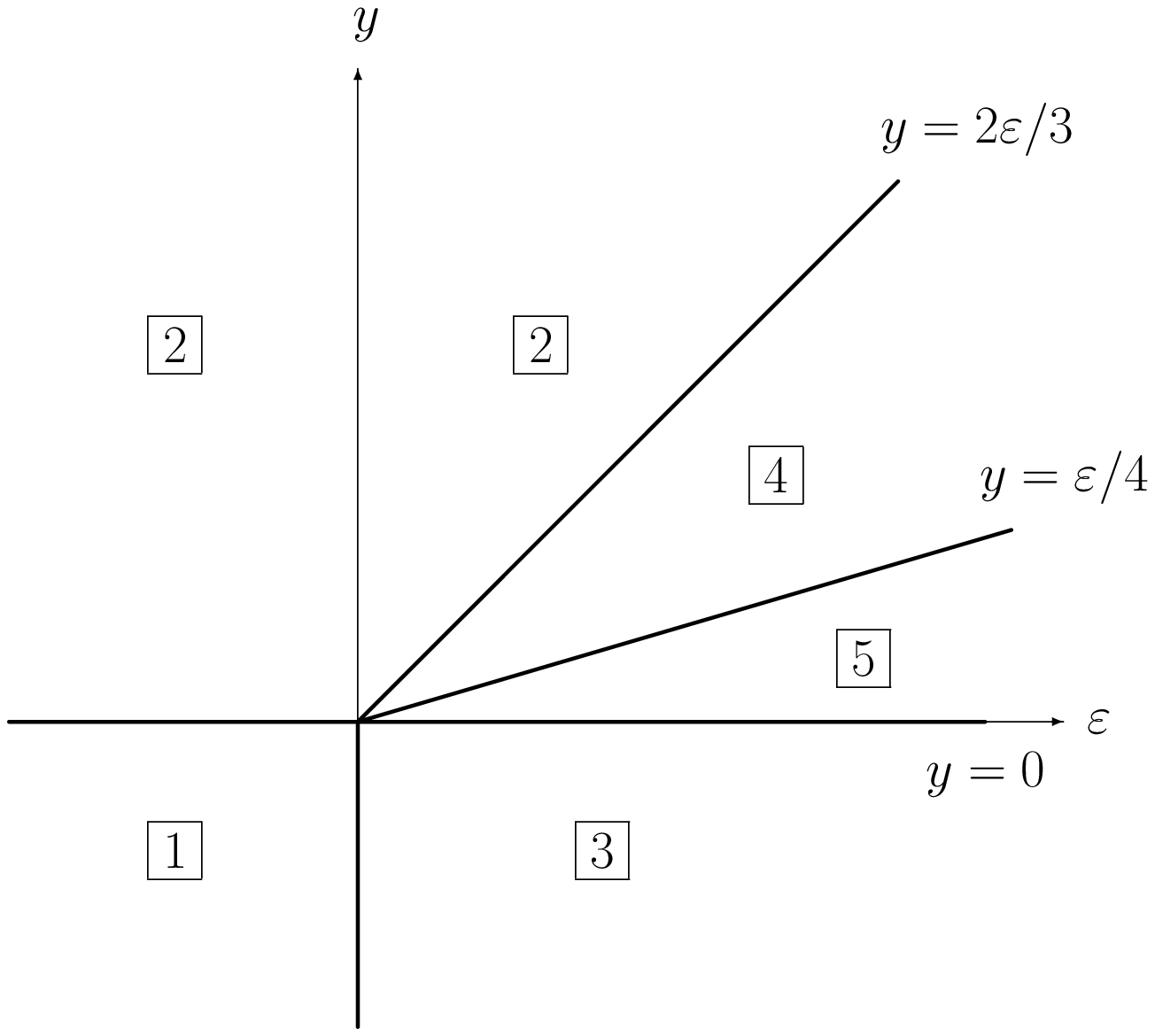}
\caption{\label{fig:patt}
Domains of IR stability of the fixed points in the models
(\protect\ref{actionF}). The numbers in boxes correspond to the fixed
points (1)--(5) in the text. For the model {\it A}, the domain (5) is
absent, and the boundary between (3) and (4) is given by the ray $y=0$,
$\varepsilon>0$.}
\end{center}
\end{figure}

In figure~\ref{fig:patt} we show the domains in the $\varepsilon$--$y$ plane,
where the fixed points listed above are IR attractive. The plot corresponds
to the Gribov case; for the model {\it A}, the domain (5) is absent, and the
boundary between the domains (3) and (4) is given by the ray $y=0$,
$\varepsilon>0$. In the one-loop approximation, all the boundaries of the
domains are given by straight lines; there are neither gaps nor overlaps
between the domains. Due to higher-order corrections to the functions
(\ref{betaA})--(\ref{betag}), the boundaries between the domains
(2), (4) and (5) can change and become curved. It can be argued,
however, that no gaps nor overlaps can appear between them to all orders;
cf.~\cite{AIK,Levy}. It is important here that the special cases $u=0$ or
$w=0$ of the full models are ``closed with respect to renormalization'' in
the sense that the functions $\beta_{u}$ for $w=0$ coincide with the $\beta$
functions of the Gribov model or model {\it A}, while the functions
$\beta_{w,e}$ for $u=0$ coincide with their counterparts in the passive
scalar model to all orders of the perturbation theory. It is also not
impossible that the absence
of the regime (5) for the model {\it A} is an artefact of the one-loop
approximation, and it will appear on the two-loop level due to nontrivial
contributions to the renormalization constants $Z_{1,4}$ in (\ref{ZoA}).

\section{Conclusion} \label{sec:Conc}

Effects of turbulent mixing on the critical behaviour were studied. Two
representative models of dynamical critical behaviour were considered:
the model {\it A}, which describes relaxational dynamics of a non-conserved
order parameter in an equilibrium critical system, and the strongly
nonequilibrium Gribov model, which describes spreading processes in
a reaction-diffusion system. The turbulent mixing was modelled by the
stochastic Navier-Stokes equation with random stirring force with the
prescribed correlation function $\propto \delta(t-t') p^{4-d-y}$.
The original stochastic problems can be reformulated as multiplicatively
renormalizable field theoretic models, which allows one to apply the field
theoretic RG to the analysis of their IR behaviour.
We showed that, depending on the relation between the spatial dimension
$d$ and the exponent $y$, the models exhibit different critical
regimes, associated with possible IR attractive fixed points of the RG
equations. For the both models, the most interesting point corresponds to
a new type of critical behaviour, in which the nonlinearity and turbulent
mixing are both relevant, and the critical dimensions depend on the two
parameters $d$ and $y$. Practical calculations of the dimensions and the
domains of IR stability for all the regimes were performed in the one-loop
approximation of the RG, which corresponds to the leading order of the
double expansion in $y$ and $\varepsilon=4-d$.

From the dimensions of the coupling constants (\ref{charges}) one could
expect that the full-scale regime (4) must take place when $y$ and
$\varepsilon$ are both positive, but the careful RG analysis has shown
that the domains of its IR stability is in fact much narrower: for the
Gribov model, in the one-loop level it reduces to the sector
$\varepsilon/4 < y < 2\varepsilon/3$, while for the model {\it A} one
obtains $0 < y < 2\varepsilon/3$. This effect leads to interesting physical
prediction: in contrary to what could be naively anticipated, the most
realistic spatial dimensions $d=2$ or 3 and the Kolmogorov exponent $y=4$
for the fully developed turbulence lie in the domain of IR stability of
the passive-scalar regime. For the
Gribov case this means that the spreading of the agent is completely
determined by the turbulent transfer. For the equilibrium model {\it A},
this is reminiscent of the observation made in \cite{Onuki,Beysens}
(however, for a conserved order parameter and non-random velocities)
that the critical fluctuations are suppressed by the motion of the fluid
and the behaviour of the system becomes close to the mean-field limit in
a strong shear flow; see also discussion in \cite{Chan}.

It is interesting to compare our results with those, obtained earlier in
\cite{AIK}, where the turbulence was modelled by Kraichnan's ensemble --
the time-decorrelated Gaussian velocity field with the correlator
$\propto \delta(t-t') \, p^{-d-\xi}$. It turns out, that the number and
the character of the critical regimes (free theory, passive scalar,
ordinary phase transition and the new full-scale regime) are the same for
the both ensembles. (For the Gribov case, the single passive-scalar regime
for Kraichnan's ensemble corresponds to the set of two regimes (3) and (5)
in the Navier-Stokes model.) What is more, in the one-loop approximations
the domains of IR stability in the $y$--$\varepsilon$ plane and the explicit
expressions for the critical dimensions coincide for the two ensembles.
(To compare the results for the two different ensembles, one has to identify
$y=3\xi$, because $\xi=4/3$ for Kraichnan's ensemble and $y=4$ for the
Navier-Stokes case correspond to Kolmogorov's velocity spectrum.)  Such
agreement allows one to conclude that Kraichnan's  ensemble, in spite of
its relative simplicity, may serve as acceptable model of turbulent mixing.

\section*{Acknowledgments}

The authors are indebted to Loran Adzhemyan, Michal Hnatich, Juha Honkonen
and Mikhail Nalimov for helpful discussions.
The authors thank the Organizers of the Third International Conference
``Models in Quantum Field Theory'' dedicated to A.\,N.~Vasiliev's
70-th Anniversary for the possibility to present the results of this work.
The work was supported in part by the RNP grant No~2.1.1/1575 and the RFFI
grant No~08-02-00125a. A.V.M. was also supported by the Dynasty Foundation.

\end{document}